# Dormant dwarf spheroidal galaxies, deactivated by Type Ia supernovae

A. Burkert[1] and P. Ruiz-Lapuente[2,3]

[1]Max–Planck–Institut für Astronomie, Königstuhl 17, D-69117 Heidelberg, Germany. E–mail: burkert@mpia-hd.mpg.de

[2]Max–Planck–Institut für Astrophysik, Karl–Schwarzschild–Strasse 1, D–85740 Garching, Germany. E–mail: pilar@MPA–Garching.MPG.DE

[3]Department of Astronomy, University of Barcelona, Martí i Franqués 1, E–08028 Barcelona, Spain. E–mail: pilar@mizar.am.ub.es

## ABSTRACT

Some dwarf spheroidal galaxies have experienced many Gyrs long periods without star formation. We show that Type Ia supernovae, formed from a first generation of stars, can delay a second epoch of star formation by several Gyrs, if the total gas mass in the dwarf spheroidals is smaller than $10^8 M_\odot$. The details depend on the assumed Type Ia supernova model and the mass fraction of stars formed in the first population. This scenario requires that dwarf spheroidals be confined by dark matter halos, as otherwise the heated gas would have been lost due to tidal effects and galactic winds. If Type Ia supernovae played an important role, heating the gas in dwarf spheroidals, they should also have increased their iron abundance during the epoch without star formation. The subsequent stellar generation should then start with a higher [Fe/H] than that of the most iron-rich stars found in the previous generation.

*Subject headings:* galaxies: dwarf spheroidals — supernovae: general — stars: white dwarfs

## 1. Introduction

The color-magnitude diagrams of dwarf spheroidal galaxies (dSph) have revealed an unexpectedly complex star formation history. The Carina dSph, for example, experienced an early epoch of star formation 11 to 13 Gyrs ago, followed by a period without detectable star formation activity. A second star formation epoch started about 5 to 8 Gyrs ago (Mighell & Butcher 1992, Smecker-Hane et al. 1996). Several Gyrs long periods of inactivity have also been found in other satellite systems of the Milky Way, like Carina (Mighell 1990) or the recently discovered dwarf spheroidal galaxy in Sagittarius (Ibata et al. 1994, Mateo et al. 1995).

These observations are quite puzzling from a theoretical point of view as all timescales, associated with processes that are considered to regulate star formation are smaller than $10^9$ yrs. For example, a few $10^7$ years after the end of the first star formation epoch heating by high-mass stars and Type II supernovae stopped. If the hot gas was still gravitationally bound, it could cool and fall back into the inner regions of the galaxy. The cooling and dynamical contraction timescales are of order $10^7$ yrs to $10^8$ yrs. One therefore would expect that new molecular clouds and another generation of stars could form a few $10^8$ yrs after the end of the previous star formation epoch.

Star formation could have been delayed for a longer time, if the gas had enough angular momentum in order to form a rotationally supported HI-disk. For HI-column densities below a critical value, the disk could remain gravitationally stable for quite a long time and would not have been able to fragment into dense molecular clouds (Gallagher & Hunter 1984, Wang & Silk 1994). The stellar components of the dSphs however do not show enough rotation in order to support such a scenario.

As the dSphs are tidally limited satellite systems it is also very unlikely that the subsequent star formation epochs were caused by late infall of new gas, several Gyrs after star formation had stopped in the systems. One rather would expect that the systems lost part of their interstellar medium as a result of tidal stripping, instead of accreting gas which was originally beyond their tidal radii.

In this *Letter* we propose a new scenario, which could explain the long periods of dormancy in dSphs: heating by Type Ia supernovae (SNe Ia). SNe Ia are considered to result from explosions of white dwarfs (WDs), which occur several $10^8$ to $10^9$ yrs after the formation of their progenitors. If the rate of explosions was high enough, radiative cooling of the hot interstellar medium could be compensated by the energy input through SNe Ia. No giant molecular clouds and stars could form for a long time. SNe Ia could then suppress star formation for several Gyrs.

## 2. Type Ia supernova scenarios

SNe Ia are assumed to result from the explosions of CO WDs in close binary systems. Different scenarios have been proposed.

In the Double Degenerate (DD) scenario (Iben & Tutukov 1984), two CO WDs merge by the emission of gravitational wave radiation. The explosion occurs due to C ignition at the center, when the more massive WD, after tidal disruption of the less massive WD, has accreted enough material to reach the Chandrasekhar mass.

In the case of symbiotic systems (SS) (Munari & Renzini 1992, Kenyon et al. 1993) a CO WD accretes material from the low-velocity wind of a red giant. The WD can explode, with a mass still lower than the Chandrasekhar mass, when C detonation is induced by the previous detonation of He, accumulated at the surface from burning of previously accreted hydrogen.

Another class of candidate systems are cataclysmic-like systems (CLS), where a CO WD accretes hydrogen on a thermal timescale from a Roche lobe filling main-sequence or subgiant companion (Iben & Tutukov 1984; Canal, Ruiz-Lapuente & Burkert 1996). Those systems are also known as "H-Algols" (Branch et al. 1995).

Figure 1 shows the expected time evolution of the SN Ia rates $\nu_{SNIa}$, normalized to the total mass of the stellar system, for the DD, SS and CLS scenarios. It is assumed that a stellar system with mass $M_*$ formed, with a constant star formation rate ($SFR = M_*/\tau_{SF}$) and a star formation timescale $\tau_{SF}$ of $10^9$ yrs. Varying $\tau_{SF}$ between $5 \times 10^8$ yrs and $2 \times 10^9$ yrs does not change the normalized SN Ia rates significantly. For the DD and CLS scenarios, the SN Ia rates depend on the common envelope parameter $\alpha$, a quantity of order unity, which measures the efficiency by which orbital energy is deposited into the ejection of a common envelope (Iben & Tutukov 1984). We show rates for $\alpha = 1$ and $\alpha = 0.3$. The rate for the SS systems has been computed, adopting an efficiency $\eta_{ss} = 1$ in producing explosions (Ruiz-Lapuente, Burkert & Canal 1995) as an upper limit and lowering this efficiency down to $\eta_{ss} = 0.1$. Note that in this scenario $\nu_{SNIa}$ scales linearly with $\eta_{ss}$.

## 3. The dwarf galaxy model

More than 10 Gyrs ago the initially completely gaseous proto-dSph experienced its first star formation epoch. Observations indicate that this event lasted of order $10^9 - 2 \times 10^9$ yrs (for a summary see e.g. van den Bergh, 1994). During this epoch a stellar system with a mass of 20 % to 60 % of the total final stellar mass of the dSph formed, that is with a mass in between $2 \times 10^5 M_\odot$ and $10^6 M_\odot$. Its high-mass stars ionized and destroyed the molecular gas clouds, forming a hot gaseous component.

Dwarf spheroidals have low metallicities, which indicates that these systems must have lost a large fraction of their metals and gas due to tidal stripping and/or through galactic winds. For example, in order to produce a stellar system with $[Fe/H] \approx -2$ only 1 % of the gas could turn into stars in the case of instantaneous recycling and efficient mixing of metals into the different gas phases, with the rest being blown out of the system. In this case, the dSphs originally contained gas with a much larger total mass, of order $2 \times 10^7 M_\odot$ to $10^8 M_\odot$. Note, that the required gas mass was smaller in the likely event of inefficient mixing of metals between the hot and cold gas phases (Burkert et al. 1993; Vader 1986, 1987). As it is preferentially the hot, metal-rich gas phase which is lost in galactic winds, a large fraction of the metals could have been lost without losing much of the gas. In this case, the total mass of the gas was more likely of order $5 \times 10^6 M_\odot$ to $10^7 M_\odot$.

At the end of the star formation event a hot gaseous bubble had formed. If heating would have been too efficient, the galaxy would have lost all of its gas and no second star formation epoch could have occured. In the cases, we are interested in, some fraction of the hot interstellar medium remained gravitationally bound. DSphs have large mass-to-light ratios, indicating the presence of a surrounding dark matter (DM) halo. The DM potential $\Phi_{DM}$ could have played an important role in keeping the gas bound.

$\Phi_{DM}$ is determined, using recent observations, which indicate that the DM halos around dwarf galaxies represent a one-parameter family with a universal density profile $\rho_{DM}$, which can be well fitted by the following formula (Burkert 1995):

$$\rho_{DM}(r) = \frac{\rho_{DM}(0) r_0^3}{(r + r_0)(r^2 + r_0^2)}. \tag{1}$$

The central DM density $\rho_{DM}(0)$ and the DM scale radius $r_0$ are functions of the total DM mass $M_{DM}$, which lies inside $r_0$:

$$\rho_{DM}(0) = 7.91 \times 10^{-2} \left(\frac{M_{DM}}{10^7 M_\odot}\right)^{-2/7} M_\odot \; pc^{-3} \tag{2}$$

$$r_0 = 0.43 \left(\frac{M_{DM}}{10^7 M_\odot}\right)^{3/7} kpc.$$

Note that for typical values of $M_{DM}$ in the range of $10^8 M_\odot$ to $10^{10} M_\odot$, the dynamical timescales for radii $r \leq r_0$ are in the range of $6.2 \times 10^7$ yrs to $1.2 \times 10^8$ yrs, respectively. Due to their high mass-to-light ratios and the fact that the hot gas was in an extended, low-density state we can also assume that the total gravitational potential $\Phi(r)$ of the dSph was dominated by its DM component: $\Phi(r) \approx \Phi_{DM}(r)$. $\Phi$ can then be determined from equation 1, using Poisson's equation.

After the end of the first star formation epoch, heating by high-mass stars and Type II supernovae stopped and the bound, hot gas fraction began to cool and fall back into the central regions again, leading eventually to a second epoch of star formation. If, however, by that time the SN Ia rate had increased sufficiently, in order to prevent the formation of cold molecular clouds, the gas would have settled into a hot, hydrostatic equilibrium state, where the rate of energy loss through cooling was balanced by the rate of energy input through SNe Ia. The next epoch of star formation would have then been delayed by several Gyrs.

The real situation could have been much more complex, given the fact that most SNe Ia exploded in the inner region, where star formation had occured initially. Heating of the inner region of a gas sphere by SNe Ia could induce large-scale turbulent motions, with hot gas moving outwards and cooler gas falling back into the center. A detailed 3-dimensional, hydrodynamical simulation of this process is beyond the scope of the present paper. As a first approximation, we neglect radial temperature gradients and turbulent flows and assume that the average gas temperature $T$ changed only on long timescales due to the variation in the balance between heating and cooling.

Given $T$ and assuming no significant radial temperature gradient, the gas density

distribution can be determined by integrating the hydrostatic equation:

$$\rho_g(r) = \rho_g(0) \times \exp\left(\frac{\Phi(r=0) - \Phi(r)}{c^2}\right) \qquad (3)$$

where $c = \sqrt{R_g T/\mu}$ is the sound velocity of the gas, $R_g$ is the gas constant and $\mu = 0.64$ is the mean molecular weight for a hot, low-metallicity gas. The central gas density $\rho_g(0)$ is determined by the known total gas mass $M_g$.

There exists a critical temperature $T_{crit}$ beyond which the gas is not confined to the dark matter halo. In this case, equation 3 has no solution for finite $M_g$ and non-zero $\rho_g(0)$. At this critical temperature the hydrostatic assumption breaks down and the system loses its gas in a galactic wind. For DM masses $M_{DM}$ of $10^9 M_\odot$ and $10^{10} M_\odot$ the critical temperature is $2.6 \times 10^4$ K and $10^5$ K, respectively. In order for the gas sphere not to condense efficiently into molecular clouds, its temperature must exceed $6 \times 10^3$ K (see section 4). This constraint and the requirement that the gas should remain gravitationally bound leads to a lower limit for the dark matter mass $M_{DM} \approx 10^8 M_\odot$.

## 4. Equilibrium states

In thermal equilibrium the gas temperature is determined by the balance between thermal energy input $(dE/dt)_{SNIa}$ due to SN Ia explosions and radiative cooling $(dE/dt)_{cool}$. The energy input rate is

$$\left(\frac{dE}{dt}\right)_{SNIa} = \eta \ \nu_{SNIa} \ M_* \ E_{SN} \qquad (4)$$

where $\nu_{SNIa}$ is shown for the SS, DD and CLS scenarios in Figure 1, $M_*$ is the total mass of the stellar system, $\eta \approx 0.25$ (Thornton et al. 1996) is the efficiency factor with which a SN Ia explosion heats the surrounding interstellar medium and $E_{SN} \approx 10^{51}$ ergs is the total energy, released in a SN Ia explosion.

The energy loss due to radiative cooling is

$$\left(\frac{dE}{dt}\right)_{cool} = \lambda(T) \int_0^\infty n_H^2 d^3r \qquad (5)$$

where $n_H$ is the local hydrogen number density of the gas, which is determined using equation 3 and $\lambda(T) = \Lambda/n_H^2$ is the cooling rate coefficient in [ erg cm$^3$ s$^{-1}$] for a low-metallicity plasma.

$\lambda$ has a temperature dependence, which makes a thermal equilibrium state possible only for gas temperatures $T \approx 10^4$K, where $\lambda$ is fast increasing with increasing temperature. For an optically thin, low-metallicity plasma in ionization equilibrium $\lambda$ reaches a maximum $\lambda_{max} = 8 \times 10^{-21}$ erg cm$^3$ s$^{-1}$ at $T = 1.8 \times 10^4$K and decreases again for higher temperatures (Fall & Rees 1985). If energy input due to SNe Ia increases the gas temperature beyond $T_{max}$, radiative cooling becomes less efficient whereas the SN Ia rate is not affected. The gas will be heated to even higher temperatures until the critical temperature is reached and the gas is lost.

Low-metallicity gas in ionization equilibrium is very inefficient in radiative cooling below $10^4$K. It therefore seems at first as if for even very low SN Ia rates one should always expect to find a thermal equilibrium state with $T < 10^4$K. Note, however, that SNe Ia drive shocks into the surrounding interstellar medium. Shocks are also expected to form in the turbulent flow which might arise due to central SN Ia heating. Shapiro and Kang (1987) demonstrate that under these circumstances and for temperatures $T < T_{min} \approx 6 \times 10^3$K non-equilibrium radiative cooling and efficient formation of molecular hydrogen becomes important. For $T < T_{min}$ molecular clouds will form efficiently which ends the quiescent phase without star formation. According to Shapiro and Kang, the cooling coefficient at $T_{min}$ is $\lambda_{min} \approx 10^{-26}$ erg cm$^3$ s$^{-1}$.

In summary, thermal equilibrium is expected only in a limited temperature range $6 \times 10^3$K$\leq T \leq 1.8 \times 10^4$K, where the cooling coefficient rises by almost 6 orders of magnitude from $\lambda_{min}$ to $\lambda_{max}$. If the SN Ia rate were so small that a thermal equilibrium requires $\lambda < \lambda_{min}$, new molecular clouds would form. In the opposite extreme case of a very high SN Ia rate, thermal equilibrium would require $\lambda > \lambda_{max}$. Then the dwarf galaxy loses its gas. In the following we will use these constraints on $\lambda$ in order to estimate whether and how long SNe Ia could delay star formation in dSphs.

## 5. Star formation, delayed by SNe Ia

Can SNe Ia heat the gas so efficiently that it is lost? Following the arguments which were presented above, this would require

$$\lambda = \frac{\eta E_{SN} \ \nu M_*}{\int_0^\infty n_H^2 d^3r} \geq \lambda_{max} = 8 \times 10^{-21} \ erg \ cm^3 \ s^{-1}. \qquad (6)$$

Numerical integrations for DM masses $M_{DM} \leq 10^{10} M_\odot$ and gas temperatures $T \leq 7 \times 10^4$K (the critical temperature for a DM halo with $M_{DM} = 10^{10} M_\odot$) show that under these conditions $\int_0^\infty n_H^2 dr^3 \geq 2 \times 10^{46} (M_g/M_\odot)^2 cm^{-3}$. According to Figure 1, the normalized SN Ia rates for SS, DD and CLS scenarios never exceed a maximum value of $10^{-12} yr^{-1} M_\odot^{-1}$ even when the efficiency parameters $\eta_{ss}$ and $\alpha$ are varied in the expected ranges. This result is true, even for combinations of the different SN Ia scenarios. Inserting these limits into equation 6, using the constraint $M_* \leq M_g$ and assuming $\eta = 0.25$ and $E_{SN} = 10^{51} erg$ we find, that SNe Ia would only be able to produce strong galactic winds for $M_g \leq 3 \times 10^4 M_\odot$ which is far below the expected values for the total gas mass in the dSphs. We therefore conclude that *SNe Ia were never efficient enough, to expell the interstellar medium in dSphs*.

Star formation will start again as soon as $\lambda < \lambda_{min}$, with $\lambda$ being determined by equation 6. This constraint leads to a minimum SN Ia rate $\nu_{SNIa,min}$ which is required to keep the gas hot:

$$\nu_{SNIa,min} = \frac{\lambda_{min} \int_0^\infty n_H^2 dr^3}{\eta E_{SN} M_*} \tag{7}$$

The integral in equation (7) is calculated for $T = T_{min}$, using equation (3). Assuming again $E_{SN} = 10^{51}$ erg, $\eta = 0.25$ and $\lambda_{min} = 10^{-26}$ erg cm$^3$ s$^{-1}$ we find

$$\nu_{SNIa,min} = 5.16 \times 10^{-21} \, f \, \left(\frac{M_g}{M_*}\right) \left(\frac{M_g}{M_\odot}\right) \, erg \, yr^{-1} \, M_\odot^{-1}. \tag{8}$$

$f$ is only a function of the mass of the dark matter halo. It increases from $f = 0.6$ for $M_{DM} = 10^{10}$ M$_\odot$ to $f = 1.3$ for $M_{DM} = 10^9$ M$_\odot$ and $f = 1.6$ for $M_{DM} = 5 \times 10^8$ M$_\odot$. $M_g$ represents the amount of gas, which, after the previous star burst, remaines gravitationally bound. $M_*$ is the mass of the stellar system which provides the SNe Ia. As typical values for dSphs we adopt $M_{DM} = 5 \times 10^8 M_\odot - 10^9 M_\odot$, $M_* = 2 \times 10^5 M_\odot - 10^6 M_\odot$ and $M_g = 10 \times M_*$ (see section 3). Note that it is very difficult to determine the total dark matter mass in dwarf spheroidals from the observed properties of their stellar components, as the scale radius of the visible component is likely to be much smaller than the radius of the dark matter halo.

Using the rates of Figure 1 and the constraint $\nu_{SNIa} > \nu_{SNIa,min}$, the timescale $\tau$ by which star formation is delayed due to SN Ia explosions can be determined. The figures 2 show $\tau$ for the SS (Fig. 2a), the DD (Fig. 2b) and the DD+CLS (Fig. 2c) scenario. The efficiency parameters were set to $\alpha = 1$ and $\eta = 1$. The shaded regions indicate the regimes of heating timescales for different $M_g/M_*$, with $M_*$ varying between $2 \times 10^5 M_\odot$ and $10^6 M_\odot$. For this mass range, SNe Ia can indeed delay star formation by several Gyrs, if the mass fraction of gas-to-stars is smaller than 30. SNe Ia, resulting from symbiotic systems are most effective in heating dSphs. They could in principle delay star formation for more than 10 Gyrs. For the SS scenario, however, the rate $\nu_{SNIa}$ depends strongly on the efficiency $\eta_{ss}$. Decreasing $\eta_{ss}$ shifts the curves of Fig. 2a by the same factor to smaller $M_g/M_*$. This effect decreases the expected delay timescales by 4 Gyrs for $\eta_{ss} = 0.5$ and leads to inefficient SN Ia heating for $\eta_{ss} \leq 0.2$. For example, assuming $\eta_{ss} = 0.1$, a $\tau = 4$ Gyrs for $M_* = 10^6 M_\odot$ requires a very small $M_g/M_* \approx 2$.

Double degenerate systems produce SNe Ia at lower rates and delay star formation only for a few Gyrs. In addition, for $\tau > 1$ Gyr, the gas-to-star ratio must be rather small. Adopting a common envelope parameter $\alpha = 0.3$, instead of $\alpha = 1$, shifts the curves by approximately 1 Gyr to larger delay timescales.

For an initial star formation epoch, which lasted of order 1 Gyrs, cataclysmic-like systems do not produce enough SNe Ia at the end of the star formation phase, in order to prevent the gas from cooling. The systems explode more than 1 Gyr later. By this time the gas would have cooled and condensed into stars again. Combining the CLS and DD scenario, early energy input is provided by the SNe Ia from DD. After 1 Gyr, the SNe Ia from CLS begin to dominate and delay star formation further by 3 to 4 Gyrs. The DD+CLS supernovae therefore induce periods without star formation which are always of order 4 to 5 Gyrs, independent of $M_g/M_*$ or $M_*$. Increasing the period of star formation to 2 Gyrs allows the cataclysmic-like systems to evolve enough, in order to produce a high SN Ia rate at the end of the star formation epoch. In this case, the SNe Ia rate from CLS systems is always large as compared to the rate of supernovae resulting from DDs, and CLS supernovae alone can delay star formation by 4 to 5 Gyrs.

In Figure 2d the total baryonic mass $M_{tot} = M_g + M_*$ during the period of star formation inactivity is shown for systems with a period of quiescence of $\tau = 4$ Gyrs. In the extreme case of low $M_g/M_* \approx 1$, SNe Ia could delay star formation in systems with baryonic masses as high as $M_{tot} = 1.6 \times 10^8 M_\odot$

## 6. Summary and Conclusions

It has been shown that SNe Ia could suppress star formation in dSphs for several Gyrs. Our calculations lead to heating timescales which are in agreement with the observations. The details depend on the total amount of gas and stars in the system and on the adopted SN Ia scenario. If the total baryonic mass exceeds $1.6 \times 10^8 M_\odot$, heating by SNe Ia becomes inefficient for $M_g > M_*$. Only in low-mass dwarf spheroidals will the star formation history be strongly affected by SNe Ia.

During the epoch without star formation, SNe Ia continuously inject iron into the interstellar medium, increasing its iron abundance. Any subsequent generation of stars which forms from that gas should therefore start with a higher [Fe/H] than the most iron-rich stars of the population which formed prior to the period of inactivity. Of order $0.6 M_\odot$ of iron are produced per SN Ia event. A stellar system with mass $M_* = 5 \times 10^5$ could then enrich primordial gas with a total mass of $10^7 M_\odot$ over $3 \times 10^9$ yrs to $[Fe/H] = -1.4$ for the SS scenario and to $[Fe/H] = -1.9$ for the DD+CLS scenario, provided that no iron-rich gas is lost. A comparison of the observed iron abundance in the younger stellar generation with respect to the older population could provide important information about the efficiency with which SNe Ia heated and enriched the dwarf spheroidals during their long periods of dormancy.

---

## Figure Captions

*Figure 1.* The normalized SN Ia rate after the end of the star formation epoch is shown for a stellar system which experienced a constant star formation rate for $10^9$ yrs. Solid line: symbiotic systems with $\eta_{ss} = 1$, short dashed line: cataclysmic-like systems, long dashed line: double degenerate systems with $\alpha = 0.3$, dotted line: double degenerate systems with $\alpha = 1$.

*Figure 2.* The figures 2a,b,c show the period of inactivity $\tau$ as a function of the ratio of gas mass to stellar mass for the SS scenario ($\eta = 1$), the DD scenario ($\alpha = 1$) and CLS scenario, respectively. The upper solid lines correspond to a stellar mass $M_*$ of $2 \times 10^5 M_\odot$, the lower solid lines correspond to $10^6 M_\odot$. The dotted lines show the period of inactivity for a stellar mass of $5 \times 10^5 M_\odot$. Figure 2d shows the total baryonic mass $M_{tot} = M_g + M_*$ of systems with $\tau = 4$ Gyrs as a function of their gas-to-star ratio for the SS (solid line), DD+CLS (dotted line) and DD (dashed line) scenarios.

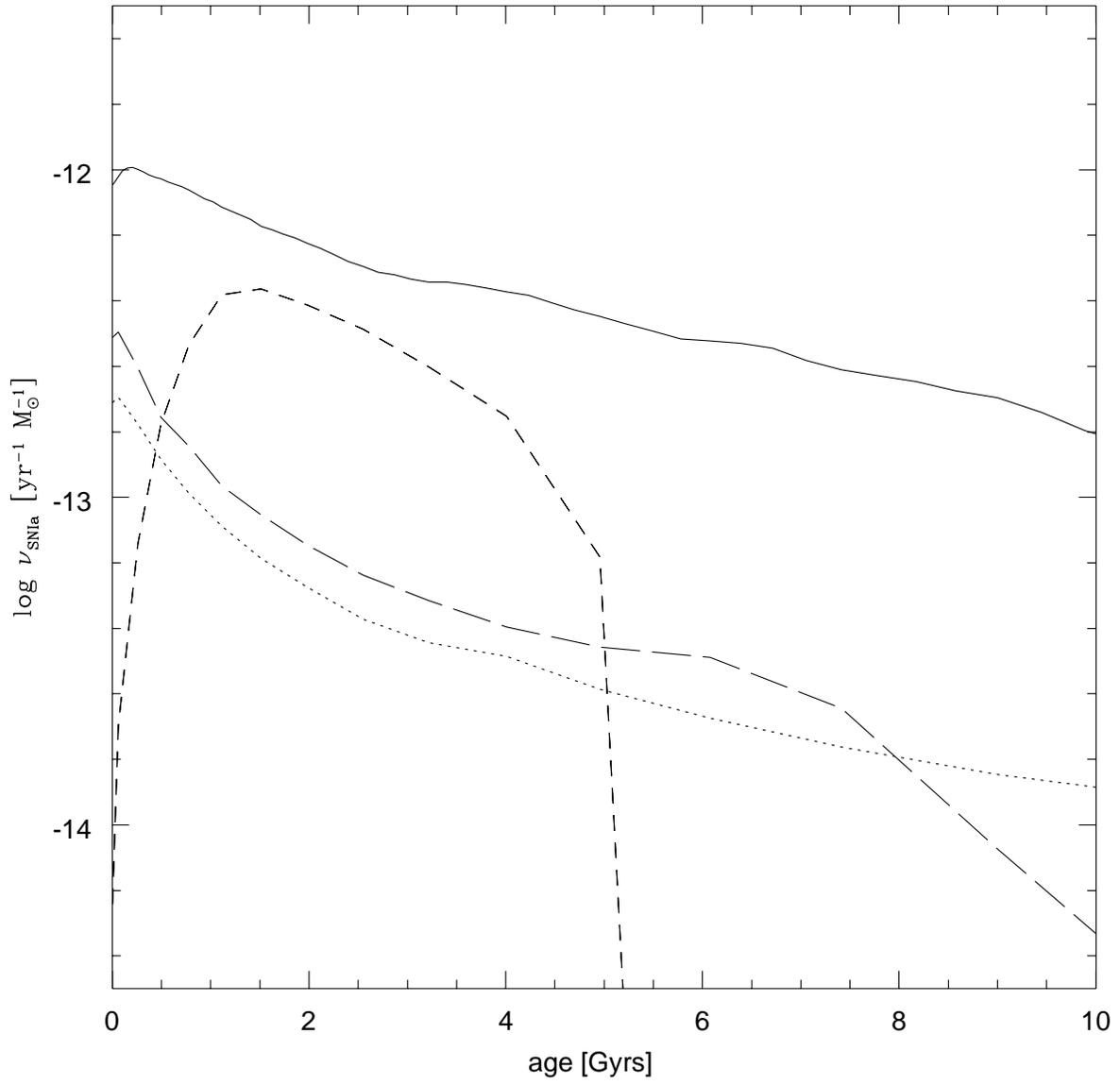

Fig. 1.—

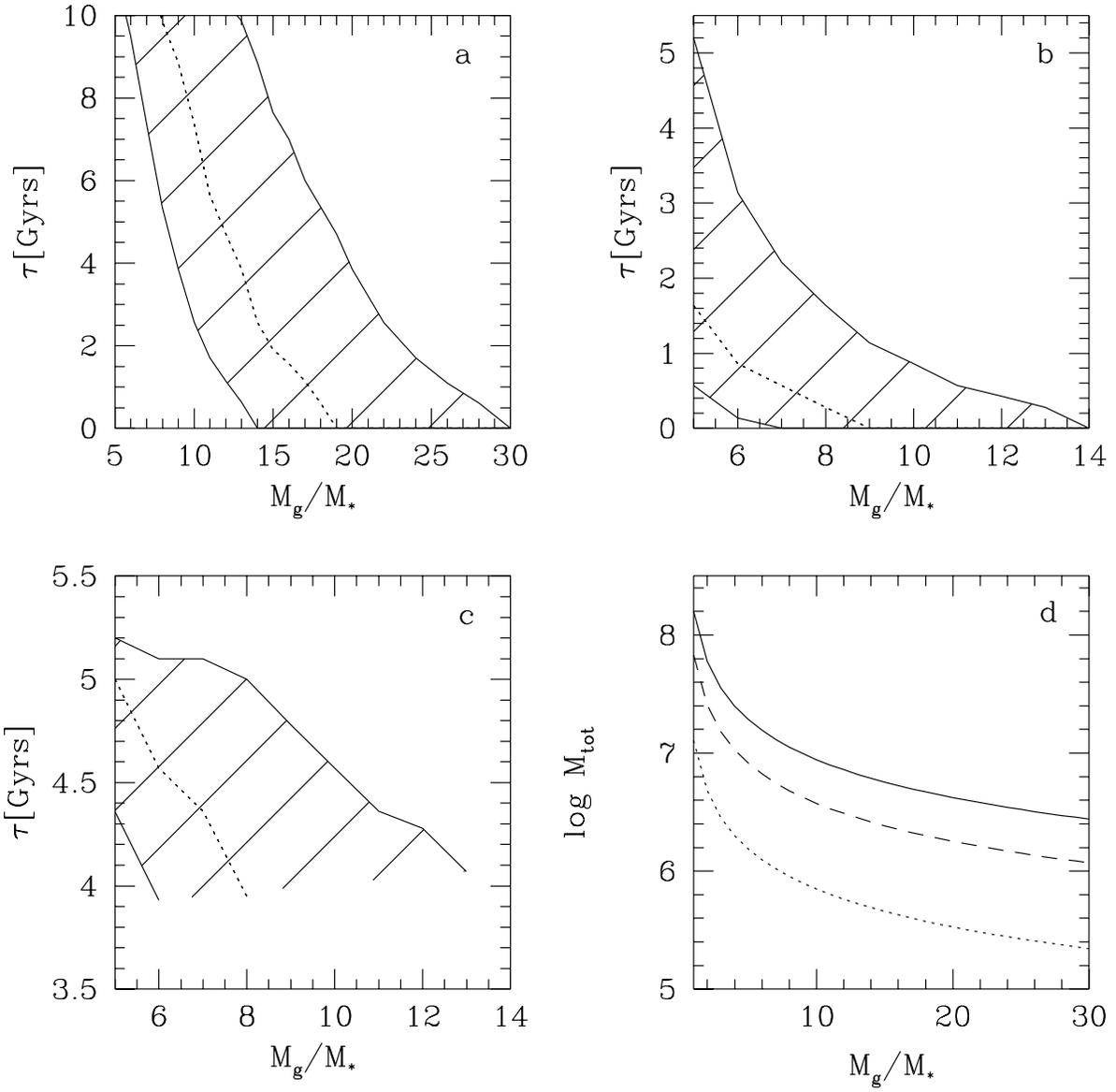

Fig. 2.—